\documentclass[a4paper, 12pt]{article} 
\usepackage[right=1in,left=1in,top=1in,bottom=1in]{geometry}
\usepackage{graphicx} 
\usepackage{braket}
\usepackage{mathrsfs}
\usepackage[T1]{fontenc} 
\usepackage{amsmath}
\usepackage{pxfonts}
\usepackage[T1]{fontenc}
\usepackage{amssymb}
\usepackage{natbib}
\usepackage{epstopdf}
\usepackage{setspace}
\usepackage{enumitem}
\usepackage{sectsty}
\usepackage{wrapfig} 
\sectionfont{\large}
\bibpunct{(}{)}{,}{a}{}{,}
\usepackage{tcolorbox}
    \usepackage{pgfplots}

\usepackage{tikz}   
\usepackage{tikz-3dplot} 

\usepackage[protrusion=true,expansion=true]{microtype} 

\newcommand{\qed}{\nobreak \ifvmode \relax \else
      \ifdim\lastskip<1.5em \hskip-\lastskip
      \hskip1.5em plus0em minus0.5em \fi \nobreak
      \vrule height0.75em width0.5em depth0.25em\fi}

\usepackage{bbm}
\usepackage{MnSymbol}
\usepackage[all]{xy}

\usepackage{xcolor,colortbl,array,amssymb}

\usepackage{epigraph}
\newcommand{\counterfactual}{%
  \mathrel{\mathop\Box}\mathrel{\mkern-2.5mu}\rightarrow
}

\makeatletter

\newlength{\bibitemsep}
\newlength{\bibparskip}
\let\oldthebibliography\thebibliography
\renewcommand\thebibliography[1]{%
  \oldthebibliography{#1}%
  \setlength{\parskip}{\bibitemsep}%
  \setlength{\itemsep}{\bibparskip}%
}

\newcommand{\x}[1]{{\color{black}#1}}

\title{\textbf{Strong Determinism}} 

\author{Eddy Keming Chen\thanks{Department of Philosophy,  University of California, San Diego, 9500 Gilman Dr, La Jolla, CA 92093-0119. Website: www.eddykemingchen.net. Email: eddykemingchen@ucsd.edu  }}

\date{\textit{Philosophers' Imprint}, penultimate version}

\begin{document}
\bibliographystyle{apa}

\maketitle 


\epigraph{What I'm really interested in is whether God could have made the world in a different way; that is, whether the necessity of logical simplicity leaves any freedom at all.}{Albert Einstein, reported by Ernst Strauss}

\begin{abstract}
A strongly deterministic theory of physics is one that permits exactly one possible history of the universe.  In the words of Penrose (1989), ``it is not just a matter of the future being determined by the past; \textit{the entire history of the universe is fixed}, according to some precise mathematical scheme, for all time.'' Such an extraordinary feature may appear unattainable in a world like ours. In this paper, I show that it can be achieved in a simple way and discuss its implications for metaphysics and philosophy of science, including natural properties, free will, explanation, and modality. First, I propose a precise definition of strong determinism. Next, I discuss its philosophical ramifications and a toy example. Finally, I provide a realistic example of a strongly deterministic (and simple) physical theory---the \textit{Everettian Wentaculus}. A surprising consequence is that whether or not our world is strongly deterministic may be empirically underdetermined. 

\end{abstract}





\nocite{albert2000time,  LeibnizPE}

\section{Introduction}

Over the last few decades, there has been significant progress in our understanding of determinism, its embodiments in concrete physical theories, and its relevance to long-standing issues in philosophy.\footnote{See for example \cite{earman1986primer, beebee2002humean, hoefer2002freedom, ismael2016physics, callender2017makes}, and \cite{loewer2020mentaculus}. See \cite{sep-determinism-causal} for an overview. } Moreover, we have seen a growing interest in super-determinism.\footnote{For recent discussions about super-determinism, see \x{\cite{hooft2014cellular}}, \cite{hossenfelder2020rethinking, ChenBell}, and \x{\cite{baas2021does}}. } In contrast, strong determinism has received little attention. In this paper, I want to examine what it is and how it impacts some of the central issues in metaphysics and philosophy of science. 


Strong determinism, according to \cite{roger1989emperor}, is ``not just a matter of the future being determined by the past; the \textit{entire history of the universe is fixed}, according to some precise mathematical scheme, for all time'' (emphasis original, p.432).   This definition, I argue, risks trivializing the distinction between determinism and strong determinism.  My first task is  to define strong determinism in terms of \textit{fundamental laws}: a strongly deterministic theory of physics is one that, according to its fundamental laws, permits exactly one nomologically possible world; our world is strongly deterministic just in case it is the only nomologically possible world. Importantly, we expect fundamental laws to be simple, which partly explains why strong determinism is difficult to achieve.  

My next task is to show that strong determinism has ramifications for a range of issues in metaphysics and philosophy of science. First, strong determinism can be regarded as a limiting case of determinism where the entire space of nomological possibilities is a singleton. A strongly deterministic theory enables an especially strong kind of explanation, as it eliminates all alternative nomological possibilities. 
Because of the physical laws, the world has to be exactly as it is.
Second,   strong determinism makes all counterfactuals (with nomologically possible antecedents) vacuously true. According to counterfactual dependence theories of causation (or modified versions in the structural equations framework), we have the surprising result that \textit{every event in spacetime causes every other event}, if strong determinism is true. It raises vexing questions about the status of causal explanations in strongly deterministic worlds. 
Third, strong determinism enables an especially strong kind of prediction; we can in principle deduce all the fundamental facts of the world (e.g. the state of the entire universe) from the fundamental laws alone, without any input from empirical observations (beyond those we need to confirm the laws). Still, strong prediction does not preclude meaningful notions of uncertainty (e.g. of self-location). 
Finally, strong determinism has implications for several debates in the metaphysics of science. For example, it vindicates a nomic version of the Principle of Sufficient Reason (PSR), sheds new light on \cite{lewis1986philosophical}'s argument for perfect naturalness, highlights the limits of \cite{loewer2020mentaculus}'s free-will compatibilism, and solves a problem in \cite{wilson2020nature}'s quantum modal realism.
  
 Is there a realistic and simple example of strong determinism? I show that the Everettian Wentaculus is such an example.  It implements strong determinism by using a deterministic dynamical law and a simple law that specifies a unique initial state. The latter is a new version of the Past Hypothesis \citep{albert2000time} in quantum mechanics.  It has the unusual feature of pinning down not just a macrostate but a microstate of the early universe. The case study of the Everettian Wentaculus has the following upshots. First, a realistic example of strong determinism may not have all the features we naively expect. Second, it may be empirically underdetermined whether our world is strongly deterministic. Third, we may have super-empirical reasons to prefer a strongly deterministic theory to a deterministic one. Finally, quantum mechanics is more hospitable to strong determinism than classical mechanics is. This stands in sharp contrast to the usual story about quantum mechanics and indeterminism.


\section{Defining Strong Determinism}

In this section, I propose a definition of strong determinism and contrast it from standard determinism and super-determinism. 

For concreteness, I first introduce the following: 
\begin{itemize}
  \item A possible world $w$: a spacetime and its material contents.\footnote{For simplicity, I assume that possible worlds have fundamental spatio-temporal structures. This is important for defining determinism but not required for defining strong determinism, which suggests that the latter is more general than the former. I will come back to this point.}
  \item The actual world $\alpha$: the actual spacetime and its material contents.
  \item Material contents: material objects and their qualitative properties.
  \item $\Omega^T$: the set of possible worlds that satisfy the  fundamental laws\footnote{In this paper, I assume there are fundamental laws and they play important roles in scientific explanations. Fundamental laws correspond to the basic principles that govern (or optimally describe) the world. In theory $T$, its fundamental laws correspond to its axioms.   Different choices of fundamental laws correspond to the axioms of different candidates for the  final theory of physics or the Theory of Everything (TOE). The fundamental laws cannot be explained in terms of deeper principles \citep[p.18]{steven1992dreams}. From them we can derive theorems of great importance and explain all  significant observable regularities. See also \cite{chenandgoldstein, chenCUP}.  Unless noted otherwise, in what follows, I use ``laws'' and ``fundamental laws'' interchangeably.
  } specified in theory $T$. 
  \item $\Omega_{\alpha}$: the set of  possible worlds that satisfy the actual fundamental laws of $\alpha$, i.e. the set of all nomologically possible worlds.\footnote{Note that $\Omega_{\alpha} = \Omega^T$ only when $T$ is the actual theory of the world, i.e. the axioms of $T$ correspond to the fundamental laws governing $\alpha$. }
\end{itemize}
I  return to the notion of fundamental laws at the end of this section. 

 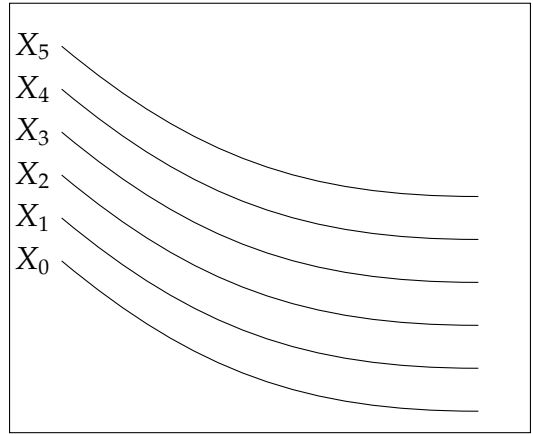
\begin{figure}
\centering	
\begin{tikzpicture}
    \begin{axis}[
        xmin=0, xmax=10, 
        ymin=0, ymax=10,
        ytick=\empty,
        xtick=\empty,
        ]
                    \draw (axis cs:1,4) to [bend right=20]  coordinate[pos=0] (dl_j) (axis cs:9,0.5); 
                            \fill (dl_j) circle (0pt) node[left] {$X_0$};
            \draw (axis cs:1,5) to [bend right=20] coordinate[pos=0] (dl_j) (axis cs:9,1.5);
                            \fill (dl_j) circle (0pt) node[left] {$X_1$};
        \draw (axis cs:1,6) to [bend right=20] coordinate[pos=0] (dl_j) (axis cs:9,2.5);
                                    \fill (dl_j) circle (0pt) node[left] {$X_2$};
                \draw (axis cs:1,7) to [bend right=20] coordinate[pos=0] (dl_j) (axis cs:9,3.5);
                            \fill (dl_j) circle (0pt) node[left] {$X_3$};
                \draw (axis cs:1,8) to [bend right=20] coordinate[pos=0] (dl_j) (axis cs:9,4.5);
                            \fill (dl_j) circle (0pt) node[left] {$X_4$};
                \draw (axis cs:1,9) to [bend right=20] coordinate[pos=0] (dl_j) (axis cs:9,5.5);
                            \fill (dl_j) circle (0pt) node[left] {$X_5$};
    \end{axis}
    \end{tikzpicture}
    \caption{Schematic illustration of a deterministic theory $T$.  $\Omega^T$ contains six nomologically possible worlds that do not cross in state space.}
\end{figure}

I define determinism as follows (see Figure 1): 

\begin{description}
  \item[Determinism$_{T}$] Theory $T$ is  \textit{deterministic} just in case, for any two $w, w'\in \Omega^T$, if $w$ and $w'$ agree at any time, they agree at all times. 
\end{description} 
  
  \begin{description}
  \item[Determinism$_{\alpha}$]    The actual world $\alpha$ is \textit{deterministic} just in case, for any two $w, w'\in \Omega_{\alpha}$, if $w$ and $w'$ agree at any time, they agree at all times. 
\end{description}
Determinism is true just in case $\alpha$ is deterministic. My definitions correspond to what \cite[p.13]{earman1986primer} calls \textit{Laplacian determinism}. For similar definitions, see \cite[pp.319-321]{MontagueFP} and \cite[p.360]{LewisNWTU}. The basic idea is that the nomologically possible worlds never cross in state space. By using the notion of a spacetime,  such definitions are more suitable for relativistic contexts as well as worlds without a fundamental direction of time.

I define strong determinism as follows (see Figure 2): 

\begin{description}
	\item[Strong Determinism$_{T}$] Theory $T$ is strongly deterministic just in case its fundamental laws are compatible with exactly one possible world,  i.e. $|\Omega^T| = 1$.
	\end{description}

\begin{description}
	\item[Strong Determinism$_{\alpha}$] The actual world $\alpha$ is strongly deterministic just in case $\Omega_{\alpha} = \{\alpha\}$. 
	\end{description}
	Strong determinism is true just in case $\alpha$ is strongly deterministic. My notion of strong determinism corresponds to the idea that the entire history of the universe is fixed by the fundamental laws of nature alone.
	
	 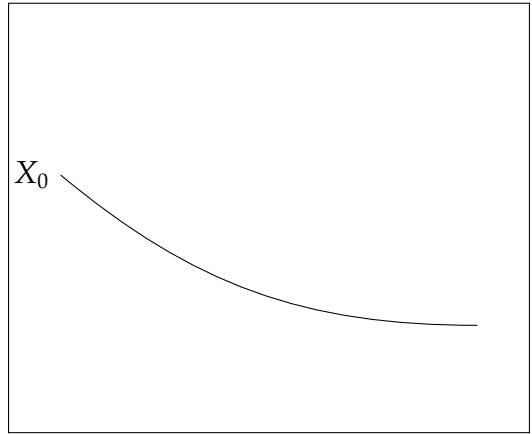
\begin{figure}
\centering	
\begin{tikzpicture}
    \begin{axis}[
        xmin=0, xmax=10, 
        ymin=0, ymax=10,
        ytick=\empty,
        xtick=\empty,
        ]
        \draw (axis cs:1,6) to [bend right=20] coordinate[pos=0] (dl_j) (axis cs:9,2.5);
                            \fill (dl_j)  node[left] {$X_0$};
        \end{axis}
    \end{tikzpicture}
    \caption{Schematic illustration of a strongly deterministic theory $T$. $\Omega^T$ contains exactly one nomologically possible world.}
\end{figure}

Under my definitions, strong determinism is stronger than determinism in a precise sense: whenever the definition of determinism is applicable, strong determinism logically implies determinism but not vice versa.\footnote{Proof: (a) Suppose strong determinism is true. Then $\alpha$ is strongly deterministic, i.e. $\Omega_{\alpha} = \{\alpha\}$. Trivially, for any two $w, w'\in \Omega_{\alpha}$, if $w$ and $w'$ agree at any time, they agree at all times. Therefore, determinism is true. (b) Suppose determinism is true. Consider this model: $\Omega_{\alpha} = \{\alpha, \beta\}$ with $\alpha$ and $\beta$ agreeing at no times. In this model, strong determinism is false because  $\Omega_{\alpha} \neq \{\alpha\}$. $\blacksquare$} \x{Strong determinism can be achieved by supplementing a deterministic theory with a boundary condition law that specifies a unique microstate of the universe at some time. See the end of this section for an example. However, not all boundary condition laws would do. A boundary condition law that specifies only the macrostate of the universe would be insufficient (e.g. the Past Hypothesis discussed in \S5.1).  Moreover, not all cases of strong determinism are achieved by supplementing a deterministic theory with a boundary condition law (\S4). }

Strong determinism is also more general than determinism. There are circumstances where strong determinism applies but determinism does not (at least not naturally). Defining strong determinism only requires the minimal notion of the cardinality of the set of models, while defining determinism requires a notion of temporal agreement, which is not always guaranteed. For example, there may be worlds without fundamental spatio-temporal structure (such as those without metrical or topological time), for which there may not be natural extension of determinism. We may not be able to say whether such worlds are deterministic, for the concept simply may not apply. But even if $w$ is such a world, we can still assess the cardinality of $\Omega_w$, the set of models compatible with the fundamental laws that govern $w$.  $|\Omega_w|$ is either 1 or larger than 1. Hence, the earlier result is valid only when the proviso holds---whenever determinism is applicable.

Let us now contrast strong determinism with super-determinism, a concept that has recently gained renewed interest in quantum foundations. \x{A super-deterministic theory is one that violates statistical independence.\footnote{\x{According to \cite{hossenfelder2020rethinking}, a super-deterministic theory also needs to be a deterministic one that is deterministic, Psi-epistemic, and local. But I shall focus on statistical  independence here.}} Roughly speaking, a theory violates statistical independence just in case the probability distribution of the physical variables describing the observed system is not independent of the detector settings. This is not a requirement of strong determinism.}
Moreover, super-determinism by itself is insufficient for strong determinism: while a strongly deterministic theory has exactly one nomologically possible world, a super-deterministic one can have (infinitely) many. Hence, strong determinism and super-determinism are logically independent.

On my view, the notion of fundamental laws is central to the definition of strong determinism (and that of determinism). \cite{roger1989emperor}, in contrast, defines strong determinism as the entire universe being fixed by some precise mathematical scheme for all time. The notion of a mathematical scheme is broader than that of fundamental laws. Although fundamental laws presumably correspond to mathematical schemes, there are many mathematical schemes that do not represent laws. Hence, Penrose's idea of strong determinism is more inclusive than the one I have. For example, as I discuss in \S5,  Penrose regards the standard Everettian theory of quantum mechanics (with a universal wave function) as an example of a strongly deterministic theory, leading to unwelcome results and risking trivializing the distinction between determinism and strong determinism. In contrast, my account does not.


There is an on-going debate about what, metaphysically speaking, fundamental laws of nature are. For concreteness, I summarize and focus on two approaches.\footnote{See \cite{sep-laws-of-nature, hildebrand2020non} and \cite{bhogal2020humeanism} for more detailed surveys.  My reason for focusing on these two is because they are two of the mostly science-friendly views in the literature, especially regarding the forms of physical laws and the direction of time. } The first is a Humean approach according to which they are merely systematizations of the material contents in spacetime:  

\begin{description}
  \item[Reformed Humeanism]  The fundamental laws are the axioms of the best system that summarizes the mosaic and optimally balances simplicity, informativeness, fit, and degree of naturalness of the properties referred to. The mosaic (spacetime and its material contents) contains only local matters of particular fact, and the mosaic is the complete collection of fundamental facts. The best system supervenes on the mosaic.\footnote{\x{A key difference between reformed Humeanism and Lewis's Humeanism \citep{LewisCounterfactuals, LewisNWTU, lewis1986philosophical} is that the latter but not the former requires fundamental laws to be regularities. See \cite[sect.2]{chenandgoldstein} and \cite[sect.2.3]{chen2018NV} for more in-depth comparisons.} }
\end{description}
 The second is an anti-Humean approach according to which laws govern and exist over and above the material contents  \citep{chenandgoldstein, chenCUP}: 
\begin{description}
  \item[Minimal Primitivism] Fundamental laws of nature are certain primitive facts about the world. There is no restriction on the form of the fundamental laws.  They govern the behavior of material objects by constraining the physical possibilities.\footnote{\x{A key difference between minimal primitivism and Maudlin's primitivism \citep{MaudlinMWP} is that the latter but not the former requires fundamental laws to be dynamical laws (in the narrow sense of being laws of temporal evolution). See \cite{chenandgoldstein} for more in-depth comparisons.} }  
\end{description}
The theoretical virtues invoked by the reformed Humean are still useful for the minimal primitivist: 
\begin{description}
  \item[Epistemic Guides] Even though theoretical virtues such as simplicity, informativeness, fit, and degree of naturalness are not metaphysically constitutive of fundamental laws, they are good epistemic guides for discovering and evaluating them. 
\end{description}
Both approaches are compatible with my definitions of determinism and strong determinism. Moreover, they are flexible regarding the form of the laws; both in-principle allow certain particular facts to be fundamental laws. For example, as I discuss in \S5, both allow the Past Hypothesis of the low-entropy boundary condition of the universe  to be regarded as a fundamental law.\footnote{\x{There are other considerations that motivate the idea that particular facts can be laws. For example, \cite{tooley1977nature} considers a law that refers to a particular physical location--Smith's garden. The initial probability distribution in Bohmian mechanics \citep{durr1992quantum} may be considered as a physical law (\cite{loewer2004david}, \cite{callender2007emergence}). \cite{hartle1996scientific, hartle1997quantum} argues that quantum cosmology requires a law of initial condition.  }}

Simplicity is important to both approaches. On reformed Humeanism, simplicity is one of the constitutive features of fundamental laws. On minimal primitivism, it is an epistemic guide for discovering and evaluating candidate fundamental laws.

 On my view,  determinism has real bite because we expect actual laws to be simple.\footnote{  \cite[pp.22-24]{russell1913notion}  and \cite[pp.22]{earman1986primer} also note that there is an important connection between determinism and simplicity, and it is mediated by the simplicity of the laws. In the end,  however, \cite[p.23]{russell1913notion} seems to reject simplicity as the solution to the trivialization of determinism, but his proposed solution in terms of uniformity of nature (more specifically, time translation invariance) can be viewed as a specific version of the simplicity requirement.} It is significant when simple laws turn out to be deterministic.  If we consider any mathematical formula regardless of its complexity,  determinism is extremely easy to achieve and can be true of any world (where the concept applies). The basic idea of determinism is that worlds never cross in state space (Figure 1). But there are infinitely many mathematical functions on state space that can meet this condition. For an extreme example, we can consider an infinitary theory $T^\infty$ whose axioms do not express simple equations. Instead, they directly specify the nomologically possible worlds of $\Omega^{T^\infty}$ (say, by giving a list of particle locations at different times) such that \textit{they never cross in state space}, rendering the theory deterministic by brute force. As long as $\alpha \in \Omega^{T^\infty}$, the theory is true and its axioms can represent the fundamental laws obtaining in the actual world.  No one bothers to write down such theories, because their axioms are in general extremely complicated and are bad candidates for fundamental laws. It is an advantage of reformed Humeanism and minimal primitivism that they recognize the importance of simplicity, either as part of the definition of what laws are or as that by which we discover or evaluate them. Hence, when characterizing determinism, it is crucial to keep simplicity in mind. Without it, determinism is easy to achieve and says almost nothing about the world, which would trivialize the distinction between determinism and indeterminism. 


Similarly,  strong determinism has real bite  because we expect actual laws to be simple.\footnote{Naturalness also plays an important role. I discuss it in \S3.3.1.}  It is even more significant when simple laws turn out to be not just deterministic but strongly deterministic. For any deterministic theory (expressed in terms of differential equations), we can always consider an extra fundamental law that stipulates the exact initial microstate of the universe. Such a new law, together with the deterministic dynamics, will make the theory strongly deterministic: given the fundamental laws (which now includes the new one), only one world is possible.  However, the axioms of such a theory will in general be extremely complicated. For example, consider a classical universe with $N$ point particles of the same mass $m$ governed by $F=ma$ with Newtonian gravitation. Add a new fundamental law specifying the complete microstate of the world at some time $t_0$, in terms of $6N$ real numbers:
\begin{equation}\label{micro}
  X (t_0) = \{\bf{q_1}, \bf{q_2}, ..., \bf{q_N}; \bf{p_1}, \bf{p_2}, ..., \bf{p_N}\}
\end{equation}
 with $\bf{q_i}$ and $\bf{p_i}$ the exact position and the momentum of the $i$-th particle in the 3-dimensional physical space.  The theory with (\ref{micro}) representing a new fundamental law will not be an attractive theory because it fails to be sufficiently simple.
 On reformed Humeanism, the specification of the exact microstate at $t_0$ will not count as an axiom in the best system of such a universe. Its gain in strength is outweighed by its cost in complexity.\footnote{\x{\cite{hall2015humean} has argued that we can make it remarkably simple by interweaving the digits into a ``phony fundamental constant.'' It deserves much more space than I can devote here.  In a realistic theory with infinite degrees of freedom, such as those of the classical fields or the quantum state, we would not be able to interweave the complete microstate of the universe into a single number.}}
 On minimal primitivism, although there is no metaphysical prohibition against such a theory, the epistemic guides tell us to look for one that better balances simplicity and informativeness.
A sufficiently simple theory that still accounts for the variety of kinds of empirical phenomena would be a marvelous achievement. We are interested in whether such theories are strongly deterministic.

\section{Consequences of Strong Determinism}

Strong determinism enables a strong type of scientific explanation and a strong type of prediction. Moreover, it  has interesting ramifications for current debates in philosophy, such as those on fundamental properties, laws of nature, free will, and modality. 

\subsection{Explanation, Causation, and Counterfactuals} 

Whether the  laws are deterministic or strongly deterministic makes a difference to the kind of explanations we obtain from a physical theory. Here, I discuss the implications for (i) strong explanations, (ii) the Principle of Sufficient Reason, and (iii) causal explanations. 

(i) \textit{Strong explanations.}  While determinism enables what I call \textit{conditional explanations}, strong determinism enables \textit{strong explanations}. For simplicity, consider again $F=ma$ with Newtonian gravitation (and appropriate boundary conditions), a familiar example of a deterministic dynamical law. Suppose it governs (or describes) a world of $N$ point particles (with positions, momenta, and Newtonian masses) moving in a 3-dimensional Euclidean space.  Its explanatory power lies in the fact that such a simple law accounts for a bewildering variety of phenomena, from falling bodies on Earth to planetary motion. For any closed system in such a world,  the law maps a state at a time uniquely to a state at another time. For the universe as a whole,  the law accounts for a general temporal pattern (cf. \cite{russell1913notion}): 
 \begin{description}
  \item[(A)] If the state of the universe  is $S$ at $t$, then the state of the universe is $S'=f(S, t, t')$ at $t'$, where $f$ is a simple function. 
\end{description}

 

We may say that the state of the universe at $t'$ is explained by the state of the universe at $t$ together with the deterministic laws. As such, the type of explanation has a conditional form: conditional on the state of the universe at $t$, deterministic laws explain the state of the universe at $t'$. Call it a \textit{conditional explanation}.\footnote{This is an instance of the Deductive-Nomological (DN) Model of explanation. See \cite{sep-scientific-explanation} for a review.} 

In contrast, strongly deterministic laws (see Figure 2) can explain more. They underwrite conditional explanations such as the above but also account for unconditional facts such as: 
 \begin{description}
  \item[(B)] The state of the universe is $S$ at $t$. 
\end{description}
The state of the universe at any time is completely explained by the laws alone. I call it a \textit{strong explanation}, in contrast to the conditional ones afforded by merely deterministic laws. 
If strong determinism is true,  every fundamental fact about the physical universe can be explained by the laws alone. If fundamental laws are where scientific explanation ultimately stops, then strong explanation may be completely satisfactory, leaving nothing unexplained. There is still the question why the fundamental laws are what they are, which I assume lies outside the scope of paradigmatic scientific explanations.\footnote{This assumption does not preclude possible metaphysical explanations of the laws. } 


In other words, if strong determinism is true, every fundamental fact about the physical universe becomes subsumed under the fundamental laws. Nothing is left to nomological contingency, chance, or randomness. 
Given the laws,  the world could not have been otherwise. To use the God metaphor: after choosing the fundamental laws, God has no more choice to make. 
If fundamental laws are where scientific explanations should ultimately rest, then strong explanation provides one of the most satisfactory explanations of the universe there can be. (Proponents of more expansive notions of scientific explanation may disagree. I discuss some of them in (iii).)

(ii) \textit{The Principle of Sufficient Reason (PSR).} There is an interesting and under-appreciated connection between strong determinism and Leibniz's PSR. Regarding determinism, \cite{sep-determinism-causal} notes that its roots lie in the PSR: 
\begin{quotation}
  The roots of the notion of determinism surely lie in a very common philosophical idea: the idea that \textit{everything can, in principle, be explained}, or that \textit{everything that is, has a sufficient reason for being and being as it is, and not otherwise}. In other words, the roots of determinism lie in what Leibniz named the Principle of Sufficient Reason.
\end{quotation}
That is a plausible suggestion.
Although there are several non-equivalent formulations of the PSR, the basic idea, as \cite{rodriguez2018principles} summarizes, is that ``there are no brute facts or truths, that is, there are no facts or truths for which no explanation can be given.'' On this characterization, strong determinism is closer to realizing PSR than determinism \x{does}. 
 As I have argued, determinism only enables conditional explanations but not strong explanations.  Even if every event can be explained by an earlier event together with the laws, mere determinism provides no explanation for \textit{why the initial event is the way it is}. Deterministic laws are (in general) compatible with many distinct initial states of the universe. Perhaps Leibniz recognizes this point when he writes:
\begin{quotation}
  For we cannot find in any of the individual things, or even in the entire collection and series of things, a sufficient reason for why they exist... [H]owever far back we might go into previous states, we will never find in those states a complete explanation [\textit{ratio}] for why, indeed, there is any world at all, and why it is the way it is.  (AG 149)
\end{quotation}
 We may formulate a nomic version of the PSR: 
\begin{description}
  \item[PSR$_{nomic}$] There is a nomic reason for every event in spacetime. 
\end{description}
Here I define  a nomic reason for an event as entailment from the fundamental laws. Determinism by itself is not sufficient for PSR$_{nomic}$, but strong determinism is.  All events in spacetime, including the initial one, are completely entailed by strongly deterministic laws. The world started in the exact initial state because, according to the strongly deterministic laws, it has to. 

The vindication of PSR$_{nomic}$ shows that strong determinism is \textit{closer} to satisfy PSR than determinism is.\footnote{To be sure, it is plausible that Leibniz has in mind a stronger version of PSR:
\begin{description}
  \item[PSR$_{nomic}^+$] There is a nomic reason for every event in spacetime, and there is a sufficient reason for the laws.  
\end{description}
Even strong determinism is not sufficient for PSR$_{nomic}^+$, as a strongly deterministic theory does not guarantee a sufficient reason for the laws. Satisfying PSR$_{nomic}^+$ requires an even stronger form of determinism, which would be interesting to explore in future work.}  Naively, we might expect that strongly deterministic laws also provide an explanation for every non-fundamental fact, such as the actual position of the table in front of me at a particular time.  However, that is not always the case, as witnessed by the Everettian Wentaculus (\S5).

(iii) \textit{Causation and counterfactuals.} Proponents of causal explanations may raise a worry. In many scientific contexts, the notion of causality is central to explanations. In philosophy of science, causality is sometimes characterized by a counterfactual dependence theory (or the related accounts in the structural equations framework).\footnote{See \cite{sep-causation-counterfactual} for an overview. There are problems of taking such an account as the analysis of causation. But even so, the counterfactual dependence theory  seems to capture an important aspect of causation for the purposes of deliberation, manipulation, and scientific modeling. } As a first approximation, we say that event $A$ is a cause for event $C$ just in case $C$ counterfactually depends on $A$:
\begin{equation}
  (A \counterfactual C) \wedge (\neg A \counterfactual \neg C)
\end{equation}
where $\counterfactual$ denotes the counterfactual conditional. The counterfactuals in such models are not counterlegals, as the causal structure should not outrun the nomic one. 

However, there is a \textit{prima facie} problem on strong determinism: there is no counterfactual possibility (the nomological state space is a singleton). Let $A(t_1)$ and $C(t_2)$ correspond to the states of the universe at any two distinct times $t_1$ and $t_2$. Assuming strong centering, $A \counterfactual C$. If strong determinism is true, on standard semantics, counterfactuals with $\neg A(t_1)$-antecedents and $\neg C(t_2)$-antecedents are vacuously true. Hence, $A(t_1)$ and $C(t_2)$ counterfactually depend on each other. This is completely general, as $t_1$ and $t_2$ can be any two distinct times. We have the surprising triviality result that every event counterfactually depends on any other event, and every event causes any other event in this world.  
The result raises vexing questions whether and how we can even make sense of causation and counterfactuals in a strongly deterministic world. It may suggest that causality has entirely disappeared in such worlds. Alternatively, it may also be interpreted as suggesting that the world is maximally causally connected. 

In my view, the compatibility of causality and strong determinism is an interesting and open question. Strong determinism is under-explored in the literature on causation, counterfactuals, and causal explanations. To invite future work on this topic, I list three options for further evaluation. 

Option 1:  We can accept the verdict of the counterfactual dependence theory but make an important revision. Causality still holds,  because the relevant counterfactuals are still true.
However,  the counterfactual dependence relations that obtain in a strongly deterministic world appear to be time symmetric, which is counterintuitive. We may prefer to recover a causal asymmetry of time. To do so, we can add a version of the Past Hypothesis (PH) as a fundamental law such that it applies to one temporal boundary of the world but not the other, and we may define a non-fundamental arrow of time as the distance away from the time that PH applies (as \cite{albert2000time, albert2015after} and \cite{LoewerCatSLaw, loewer2012two} suggest and as we do in \S5). We may then define the direction of causation as the same as the direction of time. For example,  if $A(t_1)$ and $C(t_2)$ counterfactually depend on each other but  $t_1$ is closer to the time of PH than $t_2$ is, then $A(t_1)$ causes $C(t_2)$ but not vice versa. Still, this means that any event causes any other later event.  Hence, such a world is more causally connected than we are used to. 


Option 2: We can understand the relevant counterfactuals for causal modeling and explanation as involving not the universe as a whole but the subsystems of the universe. (As \cite[pp.419-20]{pearl2009causality} acknowledges, when you describe the whole universe using interventionist models, causality disappears. See also \cite[sect. 10]{sep-causation-mani}. However, there are many identical (or similar) subsystems of the world that in which causality still exists even if it disappears at the universal level.)  Even though the universe could not have been different,  we could have been located in other subsystems of the actual universe. Hence, we may construct non-trivial state spaces for the subsystems of the universe, by using ensembles of  actual subsystems to represent counterfactual possibilities. This is not very different from a related strategy that has been explored in Everettian theories, according to which we can model counterfactual possibilities as variations in different branches of the actual multiverse. (See \cite{wilson2020nature} for a proposal.)   

Option 3: We may consider using counterlegals for causal modeling and allow causal variables to range over metaphysically possible but nomologically impossible states. This deviates from the usual practice of disallowing counterlegals, as we assume that the causal structure is fixed by the laws of nature. The relevant counterlegal possibilities can be mapped to points on phase space, configuration space, Hilbert space, and the like, which possess well-understood structure to ensure that not everything goes. (See \cite{tan2017interventions} for a related idea.)



\subsection{Prediction} 
Whether the laws are deterministic or strongly deterministic  makes a difference to the kind of predictions we obtain from a physical theory. While determinism enables what I call \textit{conditional predictions}, strong determinism enables \textit{strong predictions}.

Recall Laplace's demon:
\begin{quotation}
  We ought to regard the present state of the universe as the effect of its antecedent state and as the cause of the state that is to follow. An intelligence knowing all the forces acting in nature at a given instant, as well as the momentary positions of all things in the universe, would be able to comprehend in one single formula the motions of the largest bodies as well as the lightest atoms in the world, provided that its intellect were sufficiently powerful to subject all data to analysis; to it nothing would be uncertain, the future as well as the past would be present to its eyes.  (\cite{laplace1820}, trans. \cite{nagel1961})
\end{quotation}
If the initial value problem that Laplace has in mind is for Newtonian gravitation theory, he should have included instantaneous velocities to the things that the intelligence must know. In the best case, given the forces,  instantaneous velocities, and the positions of all particles in the universe at some time (and certain mathematical boundary conditions at infinity), a Laplacian demon can deduce all past states and all future states of the universe. However, this deduction is conditional as it requires information about the contingent state of the world at some time. In other words, in such cases, determinism enables what I call \textit{conditional predictions}:
\begin{description}  
  \item[Conditional Prediction] Conditional on the state of the universe at some time (or states of the universe at some finite interval of time), one can in-principle deduce, using the fundamental laws, the state of the universe at any time. 
\end{description}

In contrast, strong determinism enables what I call \textit{strong prediction}:
\begin{description}  
  \item[Strong Prediction] One can in-principle deduce, using the fundamental laws, the state of the universe at any time. 
\end{description}
To deduce the state of the universe at any time, one can use the fundamental laws but needs no contingent fact about the universe (beyond what one needs to confirm the fundamental laws). Such laws, if they are boundary-condition laws, may be about the state of the universe at some particular time. But if they are laws, the boundary conditions will be nomologically necessary, and we have good reasons to expect them to be simple, unlike typical microstates of the universe in a deterministic theory, which are nomologically contingent and complicated. 

Strongly deterministic laws can be predictively powerful.  To predict the outcome of the next election, merely deterministic laws are not much help as conditional prediction requires us to know the exact microstate of the universe at some time (say, the present moment). Although it is in-principle possible for us to collect all the complicated microscopic facts of the universe, it is unrealistic; our time in the universe is too short to collect enough data for such a task. Non-computability given the initial data may present additional problems for creatures like us. In contrast, strong prediction is unconditional.  We do not need to know the microstate of the universe at the present time to make the prediction. Given just the fundamental laws, we already can in principle deduce the state of the universe in, say, year 4024; if  the outcome of the next election supervenes on the state of the universe, then we can predict it with perfect accuracy.  In this sense, a Laplacian demon will have unlimited predictive power. 

However, strong prediction does not always guarantee practical usefulness for agents like us, as strong prediction does not preclude meaningful sense of uncertainty. In the case of the Everettian Wentaculus (\S5), I might have self-locating uncertainty about where I am in the multiverse. Moreover, the outcome of the next election does not simply follow from what can be strongly predicted---the state of the multiverse. I return to this issue in \S5.3.2. 


\subsection{Other Ramifications}

Strong determinism has ramifications for contemporary debates in  metaphysics and philosophy of science.  Here I discuss three examples. 

\subsubsection{Fundamental Properties: Lewis on Naturalness.}
The first example is from the landmark paper of \cite{LewisNWTU}. One of Lewis's main arguments for postulating perfectly natural (metaphysically fundamental) properties is to avoid trivializing the best-system account of laws (BSA): 
\begin{quotation}
  We face an obvious problem. Different ways to express the same content, using different vocabulary, will differ in simplicity...Given system $S$, let $F$ be a predicate that applies to all and only things at worlds where $S$ holds. Take $F$ as primitive, and axiomatise $S$ (or an equivalent thereof) by the single axiom $\forall x Fx$. If utter simplicity is so easily attained, the ideal theory may as well be as strong as possible. Simplicity and strength needn't be traded off. Then the ideal theory will include (its simple axiom will strictly imply) all truths, and \textit{a fortiori} all regularities. Then, after all, every regularity will be a law. That must be wrong. \cite[p.367]{LewisNWTU} 
\end{quotation}
 In the same paragraph, the predicate  $F$ is characterized in two different ways: 
\begin{enumerate}
  \item[(F1)] $F$ applies to all and only things at worlds where $S$ holds, which includes the actual world.
  \item[(F2)] $F$ applies to all and only things at the actual world.
\end{enumerate}
 Lewis initially defines $F$ as (F1). However, he needs the logically stronger (F2) to argue that $\forall x Fx$ strictly implies all truths.\footnote{ \cite[p.319]{loewer2007laws} and \cite[p.21]{SiderWBW} adopt the characterization in (F2).  One way to understand Lewis in this paragraph is to read the first as a general case and the second as a special case; he moves from the general case to the special one for the sake of a \textit{reductio}.  }  Given the characterization in (F2),  the deductive system $S$ axiomatized as $\{\forall x Fx\}$ is compatible with exactly one world. That is, $|\Omega^S| = 1$. Hence, $S$ is strongly deterministic on my definition. If $S$ is the best system of the actual world, strong determinism is true. 

It is a generic feature of strong determinism that all (fundamental) truths and all (fundamental) regularities about the material contents of the universe  are entailed by the fundamental laws. Consequently, on Lewis's BSA, in such a universe all such truths will be laws, albeit not all fundamental laws (which, according to \cite[p.368]{LewisNWTU}, is reserved for the \textit{axioms} of the best system). To say that this must be wrong  already presumes that strong determinism is impossible. If strong determinism is possible, the collapse of the distinction, in some worlds, between laws and mere regularities is to be expected. Hence, we need to revise Lewis's influential argument, if we accept the metaphysical possibility of strong determinism.\footnote{There are other worries about Lewis's argument. For example, his criterion for strength is logical strength and may be inappropriate for scientific theories \citep{loewer2007laws}; his sufficient condition for derived laws may be too permissive \citep{sanchezcrystallized}; his ranking method for best systems is mistaken \citep{torzaGBSA}. But I shall not focus on them here. }

I propose a revised argument, with changes italicized:
\begin{quotation}
  Given system $S$, \textit{let $F$ be a predicate that applies to all and only things at the actual world}. Take $F$ as primitive, and axiomatise $S$ (or an equivalent thereof) by the single axiom $\forall x Fx$. If utter simplicity is so easily attained, the ideal theory may as well be as strong as possible. Simplicity and strength needn't be traded off. \textit{This makes the actual world strongly deterministic, regardless of what the actual world is like. Then, after all, strong determinism is necessarily true (or true at least in all worlds where the BSA holds).} That must be wrong.
\end{quotation}
Here, the crucial premise is that strong determinism is not necessarily true.\footnote{\x{An anonymous reviewer suggests an alternative interpretation that is also reasonable: Lewis's point is that without any constraint on language, strong determinism is just trivial and says nothing about the mosaic. }} Lewis can then argue that we should postulate ``perfect naturalness'' to solve the problem.  But the revised argument clarifies that the payoff of Lewis's postulate is not to avoid the collapse of the distinction between laws and mere regularities, but to ensure that strong determinism is metaphysically contingent.

\subsubsection{Free Will: Loewer on Compatibilism}

The second example is from a recent paper by \cite{loewer2020mentaculus} on free will and determinism.  Loewer provides an ingenious reply to \cite{van1983essay}'s Consequence Argument,  based on a new theory of counterfactuals and the ``Mentaculus account'' of the temporal asymmetry of influence.\footnote{See \cite{dorr2016against} for a similar account of counterfactuals that is not explicitly based on the Mentaculus. The following discussion may also be relevant to Dorr's account but I do not have the space to discuss it here. } They are inspired by recent works in the foundations of statistical mechanics and especially the Mentaculus theory, which includes fundamental dynamical laws (such as the deterministic $F=ma$ with the force laws), a fundamental law specifying a low-entropy boundary condition (called the \textit{Past Hypothesis}), and a probability distribution over microstates compatible with the Past Hypothesis (called the \textit{Statistical Postulate}). 

  Let us focus on two key premises in Loewer's version of the Consequence Argument that he calls PAST and LAWS (where ``$\counterfactual$'' denotes the counterfactual conditional):  
\begin{description}
  \item[PAST] I have no influence over the past state at time $t$: there are no alternative decisions $d_1$ and $d_2$ possible for me at $t$ such that $d_1 \counterfactual s_1$ and $d_2\counterfactual s_2$,  where $s_1$ and $s_2$ are incompatible states of affairs that pertain to times prior to $t$. 
\end{description}

\begin{description}
  \item[LAWS] I have no influence over the laws at time $t$: there are no alternative decisions $d_1$ and $d_2$ possible for me at $t$ such that $d_1 \counterfactual  L_1$ and $d_2 \counterfactual L_2$,  where $L_1$ and $L_2$ are incompatible laws. 
\end{description}
Together with the premise of determinism and a principle that connects influence to free will, they are supposed to entail that I have no free will. As a compatibilist about free will and determinism, Loewer responds by rejecting PAST but retaining LAWS: I have (only microscopic) influence over the past of the universe but I do not have influence over the laws. It is a principled response motivated by a substantive philosophy of science (the ``Mentaculus vision''). In the context of the reply, Loewer interprets the laws as fundamental laws and the states of affairs as the (nomologically possible) microstates of the universe. 

There are many aspects of Loewer's response but I want to focus on its relevance to the theme of this paper: Loewer's compatibilism, though promising in the case of determinism, is in tension with strong determinism.  One of Loewer's insights is based on the fact that, in the Mentaculus theory,  there are distinct nomologically possible microstates compatible with the same macrostate at any time. According to Loewer, I have influence over the past state at $t$ such that, if I had done otherwise than what I actually do, the microstate of the world at time prior to $t$ would have been another microstate.\footnote{Loewer's reasoning seems to assume determinism and the condition that counterfactuals are evaluated with respect to worlds where determinism is true.} Loewer provides reasons for endorsing the other counterfactual that if I had done otherwise, the laws would not have been different. For example, even if $s_1$ and $s_2$ were incompatible microstates that counterfactually depend on my decision, they would be compatible with the same (Mentaculus) laws.   However, if the Mentaculus theory is false and strong determinism is true,  there will be exactly one nomologically possible microstate of the universe at any time. In that case, the fundamental laws are compatible with exactly one past state at any time prior to $t$. If I now had influence over the past state, I would now have influence over the laws.

One might respond by simply assuming the deterministic (but not strongly deterministic) Mentaculus theory. But the argument above is quite general as it only requires that strong determinism is metaphysically possible.   One might then try to dismiss the argument by stipulating the impossibility of strong determinism. But that misses the  point here. First, it would require some justification. (Strong determinism is possible on Loewer's package deal account of laws and properties (\citeyear{loewer2020package}).) Second, if it were justifiable, we would still have learnt something interesting: Loewer's compatibilism is compatible with determinism but incompatible with strong determinism. We may wonder: how should one generalize Loewer's compatibilism when strongly deterministic theories are allowed? When we consider (in \S5) the empirical equivalence of the Mentaculus with a strongly deterministic theory (the Everettian Wentaculus), this question becomes more urgent. 

\subsubsection{Modality: Wilson on Quantum Modal Realism}

The final example is \cite{wilson2020nature}'s quantum modal realism. Wilson proposes a bold and fascinating reintepretation of Lewis's modal realism about possible worlds in terms of Everettian (many-worlds) quantum mechanics. On the latter theory, the universal wave function gives rise to many emergent worlds.  Wilson suggests that we understand metaphysically possible worlds as Everett worlds (i.e. the decohered branches of the universal wave function that correspond to the ``emergent worlds'' discussed in \S5), and that we regard contingency as variation across such worlds. 

A natural worry is that not all contingencies are contained in the actual universal wave function. For example, even though the Schr\"odinger equation deterministically evolves an initial wave function (and by decoherence gives rise to a branching structure),  the initial wave function is nomologically contingent, i.e. not fixed by the fundamental laws. This is the case even when we impose the Past Hypothesis as a fundamental law. If nomological contingency (variation in nomologically possible worlds) is a form of contingency that Wilson aims to capture, then quantum modal realism falls short.\footnote{\x{Wilson anticipates a related worry  \cite[p.28]{wilson2020nature}.  He suggests that ``[s]ince quantum modal realists model contingency as variation across Everett worlds, there can be no contingency in an initial state that these worlds have in common.'' However, it may not be exactly the same worry, as he also discusses the issue of arbitrariness, which seems independent of the issue of contingency. }}

\x{A strongly deterministic Everettian theory can solve that problem for Wilson's proposal.} In \S5, I explain how to construct such a theory, called the \textit{Everettian Wentaculus}.  \x{In that theory, the problematic kind of contingency is eliminated.} There is exactly one nomologically possible initial condition of the Everettian multiverse and thus exactly one nomologically possible history of the multiverse. This answers the original worry about nomological contingency (though there may still be the worry for Wilson's account about how to model variations of different sets of nomological possibilities; see \cite{harding2021everettian} for a discussion). In this regard, the Everettian Wentaculus can provide a better foundation for Wilson's quantum modal realism.

These examples suggest that strong determinism can be an important resource and testing ground for philosophical theorizing. 

\section{The Mandelbrot World}
To obtain a more concrete understanding of strong determinism, let us consider a toy example, where there is a significant contrast between the complexity of the phenomena and the simplicity of the laws. The example is from the study of fractal geometry and complex dynamical systems.  Here I follow the discussion in \cite[sect. 3.2]{chenandgoldstein}. 

\begin{figure}
\centerline{\includegraphics[scale=0.4]{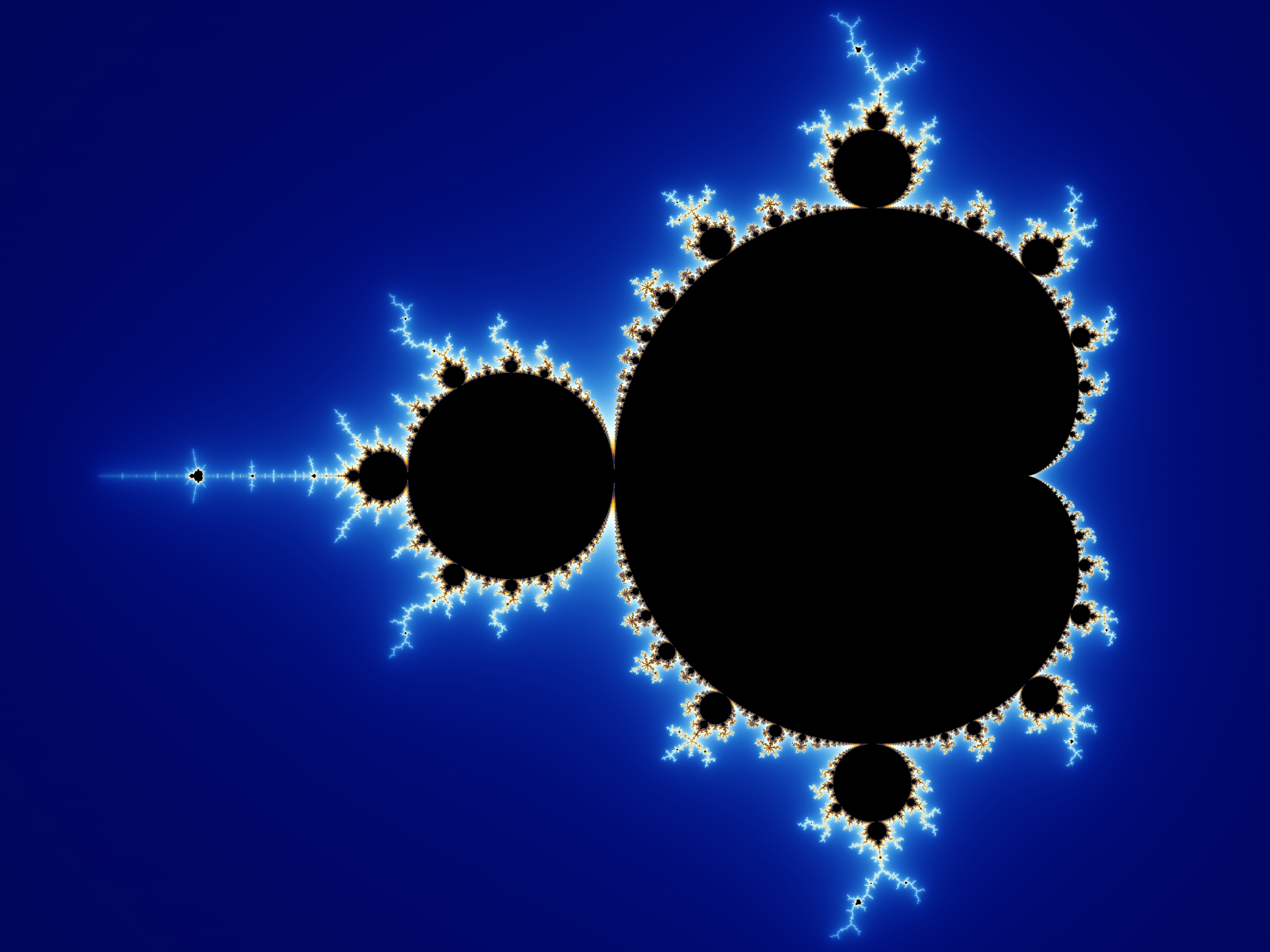}}
\caption{The Mandelbrot set with continuously colored environment. Picture created by Wolfgang Beyer with the program Ultra Fractal 3, CC BY-SA 3.0, https://creativecommons.org/licenses/by-sa/3.0, via Wikimedia Commons}
\end{figure}

Consider the Mandelbrot set in the complex plane (Figure 3), a striking example of the fractal structure, specified by the simple rule that a complex number $c$ is in the set just in case the function 
\begin{equation}\label{Mandel}
	f_c (z) = z^2 +c 
\end{equation}
does not diverge when iterated starting from $z=0$. For example, $c=-1$ is in this set but $c=1$ is not, since the sequence $(0,-1,0,-1,0,-1, ...)$ is bounded but $(0,1,2,5,26,677,458330, ...)$ is not.

Here, the pattern on the complex plane is surprisingly intricate and rich.  When we zoom in, we see sub-structures that resemble the parent structure. When we zoom in again, we see sub-sub-structures that resemble the sub-structures and the parent structure. And so on.  Interestingly, they closely resemble, but they are not exactly the same. As we zoom in, there will always be surprises waiting for us. Each scale of magnification will reveal something new.\footnote{For helpful visualizations, see \cite[ch.3]{roger1989emperor}.}  There is a puzzling pattern to be explained. 
 
Now, let us endow the Mandelbrot set with physical significance. We regard the Mandelbrot set on the complex plane as corresponding to the distribution of matter over a two-dimensional spacetime, which we call the Mandelbrot world. We stipulate that the fundamental law of the Mandelbrot world is the rule just described. The fundamental law is compatible with exactly one world.\footnote{It is worth noting that the patterns of the Mandelbrot world are not fine-tuned, as they are stable under certain changes to the law. For example, as \cite[p.94]{roger1989emperor} points out, other iterated mappings such as $	f_c (z) = z^3 + iz^2+ c $ can produce similar patterns.  } On my definition, the Mandelbrot world is strongly deterministic. 

What about explanations in the Mandelbrot world?  First, unlike the previous example, given just the pattern in the Mandelbrot world we may not expect it to be generated by any simple law.  It would be a profound discovery in that world to learn that its remarkable structure is generated by the law based on the very simple function 	$f_c (z) = z^2 +c $. The fundamental law provides a striking explanation of the pattern that leads us to say ``Aha! Now I understand.'' This also echos Penrose's emphasis on unexpected simplicity: 
\begin{quotation}
  Elegance and simplicity are certainly things that go very much together. But nevertheless it cannot be quite the whole story. I think perhaps one should say it has to do with \textit{unexpected} simplicity, where one imagines that things are going to be complicated but suddenly they turn out to be very much simpler than expected. It is not unnatural that this should be pleasing to the mind. \cite[p.268]{penrose1974role}
\end{quotation}

Second, the explanation provided by the law does not appear to be a causal or a temporal one. There is no obvious counterfactual dependence, causal or temporal ordering of events. The simple rule determines the whole world altogether. On reformed Humeanism, this simple rule is the axiom in the best system that summarizes the distribution of matter in spacetime. The axiom scientifically explains the mosaic by giving a unified account of the phenomena. On minimal primitivism, the axiom expresses the fundamental law that constrains the mosaic as a whole, even though it does not produce the mosaic moment by moment.   On both accounts, the explanation provided by the laws need not be dynamic explanations (those that unfold in time).  Unlike the example discussed in the next section, the Mandelbrot world does not have a natural structure to define a metaphysically derivative arrow of time or arrow of causation. 

What about predictions in the Mandelbrot world? As the fundamental law is non-dynamical, it does not enable the usual kind of prediction with time-evolution equations. The spacetime does not have a natural foliation into equal-time hypersurfaces, so there is no obvious notion of temporal sequences that the law acts on. Metaphorically speaking, the law treats each spacetime point individually and decides whether to place something on it. Given the law alone, a Laplacian demon can deduce everything about the world, by plugging each spacetime location (represented by a complex number) into the formula. Unfortunately, the Mandelbrot set may not be decidable in the sense of permitting a computer algorithm to calculate the exact distribution of matter in finite time \cite[p.128]{roger1989emperor}. Hence, for computationally limited creatures like us, the law may not be calculation-friendly. The calculation of the exact distribution of matter may take infinite time. (Nevertheless, since the complement of the Mandelbrot set is semi-decidable, we can in finite time obtain \textit{some} truths about the world.) 


\section{A Realistic Example}

For a realistic example of strong determinism, I turn to the Everettian Wentaculus, a novel theory of quantum mechanics in a time-asymmetric universe. It implements strong determinism by using a deterministic dynamical law and a simple boundary-condition law that specifies a unique initial microstate of the world. 

\subsection{The Everettian Wentaculus}

Let us first review the Everettian quantum mechanics (sometimes called the many-worlds interpretation).\footnote{For some recent and more detailed discussions, see \cite{wallace2012emergent} and \cite{sep-qm-manyworlds}.} In its standard formulation, it is a deterministic (but not strongly deterministic) theory that aims to solve the \textit{quantum measurement problem} and provide a consistent description of quantum phenomena. At any time, the state of the world is completely described by the universal wave function ($\Psi_t$). The time evolution of $\Psi_t$ is given by the deterministic Schr\"odinger equation: $ i\hbar \frac{\partial }{\partial t}\Psi_t = \hat{H} \Psi_t$. Fixing $\Psi_0$ suffices to fix the state of the world at any time. 

What is the quantum measurement problem that it tries to solve? Recall Schr\"odinger's famous thought experiment. If the wave function is the complete description of the system with a cat in the box, and if it always obeys the Schr\"odinger equation, then the state of the system will, after some time,  always be a superposition of the cat being alive and the cat being dead. That contradicts the assumption, suggested by observation, that after the experiment the cat is in a \x{unique} state: either alive or dead. 

Everettian quantum mechanics solves the measurement problem by embracing the \x{non-uniqueness}: yes, the cat is in a superposition of alive and dead, albeit in different ``branches.'' Many branches arise from a single universal wave function, and they correspond to different emergent worlds (which, thanks to decoherence, for all practical purposes do not interfere with each other).
  Since the observers will also experience branching, the observer in any particular branch will only observe a particular state of the cat in that branch. Everettian quantum mechanics denies that \x{the experimental outcome is unique} \textit{simpliciter}; instead, it is \x{unique} relative to a particular branch of the wave function.  On this picture, there is an emergent multiverse associated with the universal wave function. 

Let us distinguish between \textit{fundamental worlds} and \textit{emergent worlds} in Everettian quantum mechanics. Each fundamental world (whose state at a time is represented by the universal wave function) corresponds to a multiverse of (infinitely) many emergent worlds (whose states are coarse-grained descriptions of the decohered branches of the universal wave function). Fundamental worlds can be represented as curves in a state space called the Hilbert space. The theory is deterministic because those curves do not cross. The nomologically possible worlds refer to the fundamental worlds compatible with the fundamental laws.

The success of  Everettian quantum mechanics further requires solutions to two problems: (1) to provide a satisfactory ontology on which our experiences supervene, and (2) to justify the Born rule of probability in quantum mechanics. It is controversial whether the two problems have been successfully solved.  For the purpose of this paper, I set aside my doubts and grant that they have.\footnote{For relevant discussions, see \cite{barrett1999quantum, saunders2010many, ney2021world, sep-qm-manyworlds}; for a book-length defense of Everettian quantum mechanics, see \cite{wallace2012emergent}. } (As a first approximation, one may regard the Born rule probability as self-locating probability of where the agent is in the emergent multiverse. But this postulate is compatible with the determinism of the fundamental dynamics characterized by the Schr\"odinger equation.)  

Everettian quantum mechanics, as formulated, is time-symmetric in its fundamental postulates and does not yet explain the apparent (thermodynamic) temporal asymmetries, such as the melting of ice cubes, the dispersion of gas, and the diffusion of heat. There are infinitely many wave functions compatible with Everettian quantum mechanics  that do not give rise to the thermodynamic asymmetry of time.\footnote{For an overview of the thermodynamic asymmetry of time, see \citep{sep-time-thermo}. } 

To explain the asymmetries, we can adopt the Boltzmannian proposal. For concreteness, let us modify Albert and Loewer's Mentaculus theory and add the Past Hypothesis to Everettian quantum mechanics as a fundamental boundary-condition law.\footnote{ The Past Hypothesis was originally suggested in \citep{boltzmann2012lectures}[1896] and \cite{boltzmann1897} (although he seems to favor another postulate that can be called the \textit{Fluctuation Hypothesis}) and discussed in \citep{feynman2017character}[1965]. For recent discussions, see \citep{albert2000time}, \citep{goldstein2001boltzmann},  \citep{callender2004measures, sep-time-thermo}, \citep{lebowitz2008time},  \citep{north2011time},  \citep{loewer2016mentaculus},  \citep{goldstein2019gibbs}, and \cite{chen2020harvard}. The  phrase `Past Hypothesis' was coined by Albert (2000).}  In the quantum case, the Past Hypothesis is now a constraint on the macrostate realized by the initial wave function of the universe: it has low quantum Boltzmann entropy. More precisely, the initial wave function lies inside a low-dimensional subspace, denoted by $\mathscr{H}_{PH}$, the Past-Hypothesis subspace. In symbols:
\begin{equation}
  \Psi_0 \in \mathscr{H}_{PH}
\end{equation}
The size of $\mathscr{H}_{PH}$  is given by the logarithm of its dimension ($\text{log dim}\mathscr{H}_{PH}$) and its quantum Boltzmann entropy given by its size multiplied by the Boltzmann constant ($k_B$).\footnote{For an overview of Boltzmannian quantum statistical mechanics, see \cite{goldstein2019gibbs}.} The Past Hypothesis constrains the thermodynamic entropy of the world at one temporal boundary, which we might call ``the initial time.''   Given this constraint, we impose a Statistical Postulate: every wave function is equally likely as any other. (More precisely, we postulate a uniform probability distribution of wave functions compatible with $\mathscr{H}_{PH}$ with respect to the normalized surface area measure on the unit sphere in $\mathscr{H}_{PH}$.) Let us call the theory with the following fundamental laws  \emph{the Everettian Mentaculus}:
\begin{tcolorbox}
\centerline{\textbf{The Everettian Mentaculus}}
\begin{enumerate}
\item[\textbf{M1.}] The  Schr\"odinger equation. 
\item[\textbf{M2.}]  The Past Hypothesis.
\item[\textbf{M3.}] The Statistical Postulate.
\end{enumerate}
\end{tcolorbox}
On this theory, it is plausible that with high likelihood, the universal wave function will, for the overwhelming majority of times, increase in thermodynamic entropy until it reaches the maximum entropy. While Everettian quantum mechanics solves the quantum measurement problem, the Everettian Mentaculus solves, in addition, the problem of the (thermodynamic) asymmetry of time. 

The Everettian Mentaculus is deterministic but not strongly deterministic.  The Past Hypothesis constrains the initial wave function but does not pick out a unique one.  Given the fundamental laws (M1-M3), the history of the multiverse could have been different, corresponding to different choices of the initial wave function inside $\mathscr{H}_{PH}$. 

\x{Regarding the Past Hypothesis as a fundamental law leads to \textit{prima facie} issues, as it is not a dynamical law of temporal evolution and it is not time-independent.  These issues have been discussed elsewhere in the literature (see footnote \#30), and this  is a place where there can be reasonable disagreement.  Let me briefly mention some potential responses.  First, not all laws have to be dynamical laws. Recall that the Einstein equation is generally regarded as the fundamental law in general relativity, and it is not a dynamical law of temporal evolution.  Rather, it is a constraint on the entire spacetime and its matter distribution.\footnote{There are ways of converting the equation into a law of temporal evolution, but they often discard certain solutions, such as spacetimes that are not globally hyperbolic, making them physically inequivalent to the original equation.} The Past Hypothesis can be similarly regarded as a law that constrains the entire spacetime and its contents; it limits physical possibilities to only those spacetimes with the right low-entropy boundary condition. Second, not all laws need be time-independent.  Other things being equal, time-independent laws are desirable because they respect time-translation symmetry. However, whether a symmetry obtains  depends on what the world is like. It could be that empirical observations in our universe cannot be adequately captured without appealing to a law that breaks such a symmetry. As an analogy, consider an Aristotelian law that breaks spatial-translation symmetry, by privileging an absolute center of motion $C$ with respect to which all material objects must rotate. This is analogous to the Past Hypothesis that privileges a time $t_0$ when the universe must be in a low-entropy state.\footnote{\x{I thank two anonymous reviewers for discussions here. }} }
In any case,  regarding the Past Hypothesis as a fundamental law is compatible with reformed Humeanism and minimal primitivism. Just like the  Schr\"odinger equation, the Past Hypothesis can be an  axiom in the best system or a fundamental fact that constrains the behavior of fundamental objects. The reason for regarding the Past Hypothesis as simple is that the particular low-entropy boundary condition (corresponding to  $\mathscr{H}_{PH}$)  is expected to be  simple to characterize.\footnote{ A simple example of the Past Hypothesis is the Weyl curvature hypothesis: the Weyl curvature vanishes near any initial singularity \cite[p.630]{penrose1979singularities}. See \cite{ashtekar2016initial} for a generalization of Penrose's idea to loop quantum cosmology. For different types of the Past Hypothesis, including ones specified with macroscopic variables, see \cite[sect.3]{chen2018NV}. For discussions of the simplicity of the Past Hypothesis and its application to philosophy of science, see \cite[p.5]{albert2015after},  \cite[p.129]{loewer2012two}, and \cite[p.205]{callender2004measures}.} The Past Hypothesis is informative because it partly explains the thermodynamic asymmetry of time. 
 
Let us go further and construct a strongly deterministic and simple Everettian theory.  Quantum mechanics allows us to consider both pure states (represented by wave functions) and impure states (represented by density matrices). I propose a new theory, called \textit{the Everettian Wentaculus}.\footnote{The Wentaculus framework is introduced in \citep{chen2018IPH} and further developed in \citep{chen2019quantum1,  chen2018HU, chen2018valia, ChenCHETCV, chen2018NV}.} On this theory, the state of the fundamental world at $t$ is completely described by a universal density matrix ($W_t$). The time evolution of $W_t$ is given by the deterministic von Neumann equation: $i \hbar \frac{d \hat{W}(t)}{d t} = [\hat{H},  \hat{W}]$, which generalizes the Schr\"odinger equation. Fixing $W_0$ suffices to fix the state of the fundamental world at any time. Moreover, instead of postulating a uniform probability distribution over initial density matrices compatible with the Past Hypothesis subspace $\mathscr{H}_{PH}$, I postulate a particular density matrix---the natural and the canonical one corresponding to the subspace, i.e. the normalized projection. In symbols: 
\begin{equation}\label{IPH}
  \hat{W}_0 = \frac{I_{PH}}{\text{dim} \mathscr{H}_{PH}}
\end{equation}
with $I_{PH}$ the projection operator onto $\mathscr{H}_{PH}$ (the identity operator restricted to $\mathscr{H}_{PH}$).  This is called the \textit{Initial Projection Hypothesis} \citep{chen2018IPH}. It is as simple as the Past Hypothesis, as the normalized projection onto a subspace is informationally equivalent to the characterization of the subspace itself. Therefore, if the Past Hypothesis is sufficiently simple to be a fundamental law, the Initial Projection Hypothesis is too.  In contrast, on the Everettian Mentaculus, specifying an exact initial wave function requires much more information than specifying the Past Hypothesis subspace, and hence it would be a more complicated law (much like  (\ref{micro})) than the Past Hypothesis. 

I propose we regard the Initial Projection Hypothesis as a fundamental law that selects a unique initial quantum state of the universe in this new theory.
   To summarize, the Everettian Wentaculus contains two fundamental laws:
\begin{tcolorbox}
\centerline{\textbf{The Everettian Wentaculus}}
\begin{enumerate}
\item[\textbf{W1.}] The von Neumann equation.  
\item[\textbf{W2.}]  The Initial Projection Hypothesis.
\end{enumerate}
\end{tcolorbox}
The Everettian Wentaculus is strongly deterministic, since it is compatible with exactly one fundamental world. 
     \x{Given} the fundamental laws (W1\&W2), the history of the multiverse has to be what it is; it could not have been different, on pain of violating either W1 or W2. Even though the density matrix is often regarded as denoting our ignorance of the underlying pure state, on the proposed theory the density matrix plays a different role. It represents the objective and fundamental state of the fundamental world; it gives rise to an emergent multiverse.  We may say that the multiverse of the Everettian Wentaculus has ``more branches'' than that of the Everettian Mentaculus. The former has all the branches of the latter and more. Speaking loosely, the former contains not just the quantum mechanical branches but also the statistical mechanical ones. 
Hence, the fundamental world of the Everettian Wentaculus may be viewed as ontologically more expansive than that of the Everettian Mentaculus. 

If Everettian Wentaculus correctly describes the fundamental laws, the actual fundamental world will be nomologically necessary. There will be no fundamental nomic contingency or possibility beyond the actual fundamental world. If notions of contingency, chance, probability, and counterfactual make sense in this world, they have to be emergent at the level of branches and subsystems in the multiverse.  This is a proposal that completely eliminates the Statistical Postulate in fundamental physics.

\subsection{Worry: Too Easy?}

At this point, one might naturally wonder: exactly what has been achieved? It seems too easy, so there must be something wrong. Strong determinism is obtained by replacing a set of choices (initial conditions) with exactly one choice. If that is all, can't we do it much more easily in the Everettian Mentaculus, by just stipulating a particular initial microstate $\Psi_0$, thereby fixing the entire history of the multiverse?  More generally, for any deterministic theory,  can't we just stipulate exactly what the initial microstate has to be and obtain a strongly deterministic theory? Does that mean every deterministic theory is (or at least can be) strongly deterministic?

Thinking through these worries can help us appreciate what has been achieved. What sets the Everettian Wentaculus apart is the simplicity of its fundamental laws. It is a surprising discovery that our empirical experiences can be adequately described by a strongly deterministic and \textit{simple} theory, a result that would be new to many philosophers of science. 

I already discussed an example of a strongly deterministic classical mechanics at the end of \S2. The diagnosis  was that the additional postulate  (\ref{micro}) would be too complicated to be a good candidate for a fundamental law. For the standard Everettian quantum mechanics and the Everettian Mentaculus, the diagnosis is the same. We can consider an additional postulate that specifies the exact wave function $\Psi_0$ at $t_0$. Such a postulate will in general be as complicated as (if not more complicated than) its counterpart in the classical universe. The theory will no longer be an attractive one with simple axioms (expressing either the best summary or the minimal primitivist laws that govern the quantum world).  

It is non-trivial to find a strongly deterministic and \textit{simple} theory. In the Everettian Wentaculus, given the Initial Projection Hypothesis, we have a simple boundary-condition law  that specifies  a unique microstate of the fundamental world. Given also  the von Neumann equation, we have a theory that allows exactly one nomologically possible history.
\x{In} contrast, the Bohmian Wentaculus, with W1-2 plus a density-matrix version of the guidance equation, is not strongly deterministic. In the Bohmian theory, the quantum state is not everything;  the initial particle configuration is not pinned down by the Initial Projection Hypothesis. In the Mentaculus theory, the Past Hypothesis is a simple boundary-condition law but is compatible with infinitely many microstates.\footnote{The No-Boundary proposal of \cite{hartle1983wave} pins down a unique wave function of the universe. It may be another candidate of strong determinism \citep{ChenNature2023}.}

 
Recall that Penrose defines strong determinism in terms of a ``mathematical scheme'' while I define it in terms of fundamental laws. The difference manifests in our different verdicts regarding the standard Everettian theory. Penrose writes: 
\begin{quotation}
  As a variant of strong determinism, one might consider the many-worlds view of quantum mechanics (cf. Chapter 6, p.381). According to this, it would not be a \textit{single} individual universe-history that would be fixed by a precise mathematical scheme, but the totality of myriads upon myriads of `possible' universe-histories that would be so determined. Despite the unpleasant nature (at least to me) of such a scheme and the multitude of problems and inadequacies that it presents us with, it cannot be ruled out as a possibility. \cite[p.432, emphasis original]{roger1989emperor}
\end{quotation}
 On Penrose's view, standard Everettian quantum mechanics already is strongly deterministic, presumably because the actual universal wave function and the Schr\"odinger equation suffice as a mathematical scheme that fixes the history of the fundamental world. In my view, that is problematic for two reasons. First, it seems to trivialize strong determinism, rendering it a suitable target of the worry described earlier.  After all, already in classical mechanics, the world history is fixed by the dynamical laws and the initial classical microstate (\ref{micro}). It seems that any deterministic theory (whose dynamical laws are expressed as differential equations) contains a mathematical scheme that fixes the world history and is therefore strongly deterministic. Second, this view is in conflict with the usual interpretation that  standard Everettian quantum mechanics  allows many different (nomologically possible) initial wave functions. The reason we make different verdicts regarding the standard Everettian theory is because of our different definitions of strong determinism.  I do not know how to precisify the notion of ``a precise mathematical scheme'' in a way that avoids trivializing strong determinism. For that reason, I think it is better to define strong determinism in terms of fundamental laws.

\subsection{Consequences}

Let us examine the consequences of strong determinism in light of Everettian Wentaculus. 

\subsubsection{Explanation, Causation, and Counterfactuals} 

What do strong explanations look like on the Everettian Wentaculus? On this physical theory, there is exactly one nomologically possible fundamental world---the actual one. Given the fundamental laws, the world (multiverse) has to be how it is.\footnote{In the sense of metaphysical possibility, the fundamental laws could have been different. But given the earlier assumptions (\S2-3), that does not raise an additional puzzle for scientific explanations.  At the scientific level, we start from the fundamental laws and do not try to explain them further. If some laws can indeed be explained in terms of other physical theories, then that is evidence the laws are not yet fundamental. }  Hence, at the scientific level, the entire history of the multiverse is strongly explained by the Everettian Wentaculus.  This may be the ideal kind of scientific explanation on reformed Humeanism and on minimal primitivism. It also satisfies PSR$_{nomic}$. 

As discussed in \S3.1, this seems to be in tension with certain conceptions of causation and counterfactuals.  Such notions are often explicated by appeal to alternative possibilities: to understand causal relationships and counterfactual dependences, we appeal to what the actual world could have been. The Everettian Wentaculus tells us that, at the fundamental level, there is no alternative possibility. 

 Nevertheless, alternative possibilities may be recovered at the non-fundamental level of branches and emergent worlds. (This corresponds to Option 2 in \S3.1(iii).) If \cite{wilson2020nature} is right,   Everettian theories in general and the Everettian Wentaculus in particular have the structure to ground a non-fundamental notion of alternative possibilities, which may be sufficient to provide a meaningful notion of causation and counterfactuals for most ordinary contexts. Recall that the universal density matrix gives rise to an emergent multiverse due to decoherence.  For example, there will be branches where Suzy throws a rock and the window breaks and ones where Suzy does not throw a rock and the window does not break. If this approach can be successfully developed,  notions of counterfactuals and causation may still be accommodated even if the strongly deterministic Everettian Wentaculus is true. 



\subsubsection{Prediction}

What about predictions on the Everettian Wentaculus? Since the fundamental laws are strongly deterministic, strong prediction is available in the multiverse. Hence, a Laplacian demon can deduce the entire history of the multiverse from the laws alone, without any input about contingent matters of fact.

Whether the entire world is computable depends on whether W1 and W2 are computable. However, even granting their computability, strong prediction does not preclude meaningful notions of uncertainty for situated agents like ourselves.  In Everettian theories, since every possible outcome of each experiment is realized in some branch, there needs be an account for the Born rule probability that situated observers can use.  Everettians try to solve this problem by appealing to either decision theory or self-locating uncertainty, placing the source of such probability in the agents rather than the nomological structure of the world.\footnote{See \cite{sep-qm-manyworlds} for a survey; for an example of the decision-theoretic approach, see \cite{wallace2012emergent}; for an example of the self-locating uncertainty approach, see \cite{sebens2016self}. For generalizations of some of those solutions to density-matrix theories such as the Everettian Wentaculus, see \cite{ChenChua}. } The goal is to justify (both qualitatively and quantitatively)  the Born-rule probability so that we can make sense of how outcomes of measurement do in fact confirm Everettian quantum mechanics. I granted earlier that the probability problem(s) can be solved, otherwise Everettian quantum mechanics is \textit{already} subject to decisive refutation. Insofar as we consider Everettian quantum mechanics a live empirical hypothesis (which many people do), we have to presuppose that it makes sense to talk about Born-rule probability, either through how much I prefer certain rewards or how likely I am located in a particular branch of the multiverse. 

Hence, there is (by assumption) still a meaningful sense of uncertainty and probability. Either I will act as if I am uncertain or I will lack information about which branch I am on. Either way, prediction of the particular outcome of experiment will be effectively probabilistic, in accord with the Born rule. Therefore, even though strong prediction is available, at the level of practical action and deliberation,  predictions will remain effectively probabilistic.

This point may generalize to other quantum theories. Already in standard Bohmian mechanics, determinism of the fundamental laws is compatible with absolute uncertainty---``when a system has wave function $\psi$ we cannot know more about its configuration $X$ than what is expressed by $|\psi|^2$'' \cite[p.885]{durr1992quantum}.  Now, consider a \textit{hypothetical} Bohmian theory that implements strong determinism by postulating a simple and compelling law that picks out not just a unique initial quantum state (W2) but also \textit{a unique initial particle configuration}. As long as the simple law does not pick out an atypical configuration (displaying quantum non-equilibrium), there can be unpredictability due to absolute uncertainty and the dispersion in the dynamical equation \cite[pp.885-86]{durr1992quantum}, as strong determinism is compatible with quantum equilibrium hypothesis. Our predictions in such a world will still be effectively probabilistic, in accord with the Born rule ($|\psi|^2$). In this case, our uncertainty can be understood as that of self-location (in space and time)---we may be uncertain of which subsystem we are in.

\subsubsection{Empirical Equivalence}

Assuming that the Everettian problem(s) of probability can be solved,  Everettian theories are empirically equivalent to textbook quantum mechanics, insofar as the latter makes unambiguous predictions. Moreover, Everettian theories are empirically equivalent to their Bohmian counterparts, as both assign the same Born rule probabilities to measurement outcomes. Furthermore, spontaneous collapse theories such as GRW can be made approximately empirically equivalent to Everettian theories. Hence, the Everettian Mentaculus, the Bohmian Mentaculus, the Bohmian Wentaculus, and the Everettian Wentaculus are all empirically equivalent. They are also approximately empirically equivalent to the GRW Mentaculus and the GRW Wentaculus.\footnote{\x{The GRW  theory employs time-asymmetric laws, but  we still need to add a low-entropy boundary condition to get everything started \cite[p.161-62]{albert2000time}.   }} 

Given their empirical equivalence, in a time-asymmetric quantum world like ours (assuming unitary dynamics of the quantum state and setting aside GRW), we cannot find out whether strong determinism is true or false by experiments or observations alone.
\x{T}his is true regardless of the measurement devices we use, based on any technology present or future. The question of strong determinism will forever be empirically underdetermined.

One might respond by pointing out that our notion of \textit{interesting} strong determinism already appeals to the super-empirical virtue of simplicity. After all, any deterministic theory is empirically equivalent to an \textit{uninteresting} strongly deterministic theory that turns out to be complicated (by stipulating the exact microstate such as in (\ref{micro}) or $\Psi_0$). In response, a stronger point can be made.  We might have thought that no sufficiently simple theory can account for our empirical experiences and validate strong determinism. But that turns out to be wrong. \x{Everettian Mentaculus and Everettian Wentaculus are not only empirically equivalent, but also have equally simple laws}. Moreover, there are super-empirical considerations to regard the Wentaculus as better than the Mentaculus \citep{chen2018IPH, chen2018valia, chen2018HU, chen2018NV}. If so, we may have reasons to prefer the Everettian Wentaculus to the Everettian Mentaculus. 

This result is interesting even for people, such as myself, who do not think that the probability problem has been solved in Everettian theories. First, it shows that whether our world is strongly deterministic turns on conceptual issues of probability. Second, it is quantum mechanics but not classical mechanics that is hospitable to strong determinism, because classical mechanics does not contain an attractive theory that is strongly deterministic. This stands in sharp contrast to the traditional belief that the quantum world is more indeterministic.

\section{Conclusion}

Strong determinism holds when the actual world is the only nomologically possible world. Philosophers and physicists have reasons to be interested in strong determinism. As illustrated by the Mandelbrot World, it enables strong explanations and strong predictions. It also raises vexing questions about the status of causation and counterfactuals. The Everettian Wentaculus,  a realistic and simple strongly deterministic theory, teaches us that strong determinism may well be true but does not always have the features we naively expect. In particular, it does not preclude meaningful notions of uncertainty. 

Whether our world is strongly deterministic may not be settled empirically, if the Everettian Wentaculus is empirically equivalent to other formulations of quantum mechanics in a time-asymmetric universe. Moreover, certain super-empirical considerations may even favor the Everettian Wentaculus over its competitors. It is surprising that quantum mechanics is more hospitable to strong determinism than classical mechanics is.
Whether or not strong determinism is true, it is closer to the actual world than we have presumed, with implications for a variety of topics in philosophy and foundations of physics. 
I have examined only some of them here, and I hope this paper will serve as an invitation for others to explore this rich concept. 


\section*{Acknowledgement} 

Versions of this paper have been presented at the Lake Arrowhead Conference of the California Quantum Interpretation Network, the Quantum Foundations Seminar at the University of Surrey, UC San Diego, UC Irvine,  UC Riverside, UCLA, Johns Hopkins University, University of British Columbia, NYU Shanghai, Chinese University of Hong Kong, Chinese Academy of Science Institute of Philosophy, Tel Aviv Conference on the Many-Worlds Interpretation, and Metro Area Philosophy of Science Group. I thank the participants in those events, especially  Jim Al-Khalili, Fatema Amijee, Alisabeth Ayars, Jeff Barrett, Paul Bartha,  Craig Callender, Adam Chin, Heather Demarest, Alison Fernandes, John Martin Fischer,  David Gilbert, David Glick, Nicholas Hanson-Holtry, Daniel Alexander Herrmann, Mario Hubert, Soofia Lateef, Matthew Leifer, Hanyu Liu, Helen Meskhidze, Michael D Nelson, Alexandra Mary Newton, Alyssa Ney, Matic Kastelec, Don Page, Lauren Ross, Simon Saunders, Margaret Schabas, Sheldon Smith, Kyle Stanford, Eric Schwitzgebel, Lev Vaidman,  David Wallace, Sean Walsh, Eric Watkins, Ken Wharton,  Jingyi Wu, and Jiji Zhang. I am also grateful for helpful discussions with Emily Adlam, Eugene Chua, David Danks, Christopher Dorst, Sam Elgin, Mathias Frisch, Sheldon Goldstein, James Hartle, Marc Lange, Barry Loewer, Daniel Rubio, Alessendro Torza,  James Weatherall, and Alastair Wilson.  I thank Don Rutherford and Shelly Yiran Shi for discussions about the connections between Leibniz's PSR and determinism. I benefited immensely from the research assistance of Shelly Yiran Shi, who provided detailed comments on multiple drafts.  This work was made possible partly through the support of Grant 62210 from the John Templeton Foundation. The opinions expressed in this publication are those of the author and do not necessarily reflect the views of the John Templeton Foundation.


\bibliography{test}


\end{document}